\documentclass[12pt]{article}
\usepackage[margin=2.3cm]{geometry}
\usepackage{setspace}
\usepackage{graphicx}
\usepackage{bm}
\usepackage{mathptm,amsmath} 
\usepackage{graphicx}
\usepackage{amssymb}
\DeclareMathAlphabet{\mathcal}{OMS}{cmsy}{b}{n}

\begin{document}

\doublespacing

\title{{Physics of counterion mediated attractions between double-stranded DNAs}
\author{Fabien Paillusson}\\ \small Department of Chemistry, University of Cambridge,\\ \small Lensfield Road, CB2 1EW, Cambridge (UK) }

\maketitle

\begin{abstract}
Attraction between two indentical objects uniformly charged in solution seems intuitively surprising on the one hand because we know that such objects do repel in vacuum and on another hand because this simple mental picture is still preserved within the Poisson-Boltzmann theory of ion-mediated interactions. We will see, however, that by relaxing some contraints on the models used, one can find ion-mediated attraction between uniformly charged objects in solution. This allows to give some insights on the possible way two DNA double strands attract each other in many biologically relevant cases.
\end{abstract}


\section{Introduction}\label{Introduction_Paillusson}
Attraction between two indentical objects uniformly charged in solution seems intuitively surprising on the one hand because we know that such objects do repel in vacuum and on another hand because this simple mental picture is still preserved within the Poisson-Boltzmann (PB) theory \cite{Parsegian, Neu,Sader}. We recall that the PB theory is a mean field approximation \cite{Orland} of a primitive model of electrolytes \cite{HL00} in which ions are point like and solvent is simply a uniform medium of dielectric constant $\varepsilon$. Attraction mechanisms mediated by counterions require therefore to modify at least one of the aforementioned assumptions. We will see that one can find attraction by a) finding a situation in which two non uniformly like charged objects can attract each other even in vacuum, b) adopting another theoretical approximation for the same model of electrolyte and c) changing the model for the ions.\newline Before going on, we introduce the characteristic length scales arising when studying counterions of valency $q$ subject to thermal fluctuations of magnitude $k_B T$ next to a uniformly charged macroions with charge density $\sigma$ in a solvent of permitivity $\varepsilon$. We note $q^2 l_B \equiv q^2 {e}^2/4\pi \varepsilon k_B T$ the Bjerrum length below which electrostatic repulsion dominates the thermal motion of counterions and $\mu_G \equiv (2\pi l_B q \sigma)^{-1}$ the Gouy-Chapman length that estimates the extent of the condensation layer on a plate owing to thermal fluctuations. Finally, we note $a_{\perp}$ the typical lateral distance between two neighbouring counterions on a plate such that $\pi a_{\perp}^2 \sigma = q$.
\section{Planar Kornyshev-Leikin (KL) theory}
One possible way to get like charge attraction out of the PB theory is to find an electrostatic configuration that leads to attraction in vacuum. This is in fact possible as soon as the macroions are not uniformly charged but instead diplay, say, an assembly of quasi-dipoles. This is the idea behind the so called KL theory \cite{Kornyshev99, Kornyshev01, Kornyshev07, Kornyshev11} that was originally done considering a realistic model of the charge pattern on DNA. In what follows, we shall present a simplified version of it with a planar geometry to focus on the physics rather than the equations.
\subsection{One plate in a salt solution}
We consider a plate at $z=0$ whose surface charge distribution is composed of very thin stripes extending in the $x$ direction of negative line charge density $-|\lambda_0|$. The distribution is periodic in the $y$ direction such that there is a spacing $H$ between two stripes. Let us put this object in an electrolyte solution whith an inverse Debye length $\kappa=\sqrt{\sum_i 4\pi q_i^2 l_B n_i}$. If, moreover, the solution contains plate counterions with a high valency, then it is assumed that most of those will adsorb on the plate so as to neutralize its charge. In this section we will consider that the adsorbed counterions mostly form positive stripes parallel to the original negative ones and each one of them lies exactly in at equal distance from two neighbouring negative stripes. Their line charge density is defined as $\theta |\lambda_0|$ with $0<\theta <1$. In the end, the fixed charge density on such a ``dressed`` plate can read $\rho_L(y,z) =  \delta(z) (\sigma_d(y)+\sigma_m(y))$ with:
\begin{eqnarray}
&& \sigma_d(y) \equiv |\lambda_0|\sum_{i \in \mathbb{Z}} \left(\delta(y-iH-H/2) -\delta(y-iH) \right) \label{KL2} \\
&& \sigma_m(y) \equiv |\lambda_0|\sum_{i \in \mathbb{Z}} (\theta -1)\delta(y-iH-H/2)  \label{KL2bis}
\end{eqnarray}where we simply used the identity $\theta = \theta +1 -1$ to subdivide the surface charge density into a purely dipolar contribution (alternating stripes of exactly opposite sign) and a contribution arising only from stripes of the same sign. Both $\sigma_d(y)$ and $\sigma_m(y)$ are even and periodic functions of period $H$ and it is convenient to express them in term of their Fourier series components $\sigma_d^k$ and $\sigma_m^k$\footnote{Any even and periodic function $f(y)$ of period $H$ can be written as:
\begin{eqnarray}
 f(y) = f_0 + \sum_{i=1}^{\infty} f_k \:\cos \left(k\frac{2\pi y}{H}\right) \nonumber
\end{eqnarray}with 
\begin{eqnarray}
 f_0 \equiv \lim_{h \rightarrow 0}\frac{1}{H}\int_{-H/2+h}^{H/2+h}dy\: f(y);\:\:\:f_k \equiv \lim_{h \rightarrow 0}\frac{2}{H}\int_{-H/2+h}^{H/2+h}dy\: f(y) \cos \left(k\frac{2\pi y}{H}\right) \nonumber
\end{eqnarray}where the use of $h$ prevents problems when integrating over Dirac combs of period $H$.}. The $k=0$ harmonic mode is nothing but the average surface charge density that is trivialy $\sigma_d^0=0$ for $\sigma_d(y)$ and $\sigma_m^0=(\theta-1)|\lambda_0|/H$ for $\sigma_m(y)$. For the rest of the components, it easily found that $\sigma_d^k=2|\lambda_0|[(-1)^k-1]/H$ and $\sigma_m^k=2|\lambda_0|(\theta-1)(-1)^k/H$.  Within a KL-like theory, we next consider that the mean electrostatic field $\varphi$ satisfies the linearized PB equation \cite{Kornyshev99,Kornyshev01}. Expressing it directly it terms of its Fourier components $\varphi_k$, we get the following set of equations:
\begin{eqnarray}
\frac{d^2 \varphi_k(z)}{dz^2}-\kappa_k^2 \varphi_k(z) = -\frac{1}{\varepsilon}\delta(z)(\sigma_d^k+\sigma_m^k) \label{KL5}
\end{eqnarray}where $k \geq 0$. We notice from Eq. \eqref{KL5} that, the higher $k$, the higher the effective inverse screening length $\kappa_k = \sqrt{\kappa^2+k^2 4\pi^2/H^2}$. Since we are interested in the far field generated by the plate, we will focus on the first two modes $k=0$ and $k=1$. The electrostatic potential then reads far from the plate:
\begin{equation}
\phi(y,z) \sim \frac{|\lambda_0|(\theta-1)}{2H \varepsilon}e^{-\kappa z} - \frac{|\lambda_0|(\theta+1)}{H \varepsilon}e^{-\kappa_1 z} \cos(\frac{2\pi}{H}y) \label{KL6}
\end{equation}
\subsection{Interaction between two plates}
We now look at the interaction of the plate described in the previous section with a similar plate but shifted in the $y$ direction by an amount $\delta y \leq H/2$ and at a distance $z=L$ from the former plate. Its fixed charge density reads then $\rho_R(y,z)=\rho_L(y-\delta y,z-L)$. We also assume that the plates are far enough to neglect the entropic contribution from the ions to the interaction energy. In this case, only the electrostatic contribution coming from the far field in Eq. \eqref{KL6} matters. The corresponding energy per unit area reads then
\footnote{We define it as being:
\begin{eqnarray}
 \mathcal{U}_{el} \approx \lim_{h \rightarrow 0} \frac{1}{H}\int_{- H/2+h}^{H/2+h} dy\: \phi(y,L)(\sigma_d(y-\delta y)+\sigma_m(y-\delta y)) \nonumber
\end{eqnarray}
}:
 \begin{equation}
 \mathcal{U}_{el}(L) \approx \frac{\lambda_0^2}{2\varepsilon H^2}\left[(\theta-1)^2 e^{-\kappa L} + 2(\theta +1)^2 \cos\left( \frac{2 \pi \delta y}{H} \right)e^{-\kappa_1 L}  \right] \label{KL8}
 \end{equation}
From Eq. \eqref{KL8}, we see that as soon as the shift $\delta y > H/4$, the second term in the r.h.s. becomes attractive. This attractive interaction is maximum if the shift $\delta y$ between the the charged patterns on the plates equates exactly half the period $H$. As a matter of fact, in that case, each stripe on a plate would be facing an almost opposite stripe on the other plate thus leading to an attraction. Overall, a net attraction is all the more likely if $\theta$ is close to unity as the magnitude of the repulsion in Eq. \eqref{KL8} is proportional to the square of $(\theta-1)$.

\section{Strong Coupling Regime} \label{SC_Paillusson}
The electrostatic coupling in a charged system with counterions can be described by a unique parameter $\Xi \equiv 2 \pi q^3 l_B^2 \sigma$ so that the limit of vanishing $\Xi$ for an exact field theoretical description of the system gives the PB theory \cite{Orland}. The SC regime can then be expressed as the opposite limit when $\Xi$ goes to infinity. The following two subsections summarizes two different appoaches of this SC regime, the Virial Strong Coupling (VSC) appoach and the Wigner Strong Coupling (WSC) approach.
\subsection{Virial strong coupling}
When $\Xi$ tends to infinity, the counterions condense onto the macroions they neutralize \cite{Rouzina96, ShklovskiiPRL99, Lau01, Lauthesis,Moreira02}. At finite values of $\Xi$, few counterions are desorbed from the surfaces and the electrolyte is effectively decomposed into a condensed liquid phase on the macroions and a bulk vapor. The bigger $\Xi$, the smaller the vapor density and hence its fugacity $\lambda$. At equilibrium, the vapor fugacity has to equate that of the liquid so that the counterion fugacity tends to zero as $\Xi$ tends to infinity. R. Netz suggested accordingly a Mayer expansion-based \cite{HansenMcDonald} $1/\Xi$ expansion of the dimensionless one-particle density $\tilde{\rho}_1$ (in units of $\mu_G^{-3}$) and the free energy $\beta F$ of this system \cite{Netz01}. This theory has been discussed at length in the past \cite{Moreira01,Moreira02,Netz04,Naji04,KanducPRE08} although its capacity at capturing what is happening in the system has been questioned recently \cite{TrizacPRL11,Trizac11} and will have a dedicated chapter in this volume. For these reasons, we will simply present its single particle picture predictions below:
\begin{eqnarray}
&& \beta F(N) = \beta W_0 - N \ln \int d^3 \tilde{r} \:e^{-\phi(\mathbf{r})} + \mathcal{O}(\Xi^{-1})\label{eqPai13} \\
&& \tilde{\rho}_1(\mathbf{r}) = \frac{C}{\Xi}e^{-\phi(\mathbf{r})} + \mathcal{O}(\Xi^{-2}) \label{eqPai8} 
\end{eqnarray}where $\beta W_0$ is the energy of the system in absence of counterion, $\mathbf{\tilde{r}}\equiv \mathbf{r}/\mu_G $ is a dimensionless position vector and $\phi$ is the external potential (in units of $k_B T$) owing to the macroions. The constant $C$ in Eq. \eqref{eqPai8} can be found by requiring the integral of $q\rho_1$ to be equal to the total charge carried by the macroions.
\subsection{Wigner strong coupling}
The WSC approach starts from the fact that we know what is the exact ground state of a system of charges next to a plate (a Wigner crystal with a triangular lattice) and therefore we can expand thermodynamic quantities around this ground state \cite{Rouzina96,Shklovskii99,LauPRL2000,Lau01,Lauthesis,TrizacPRL11,Paillusson-Trizac,Trizac11,Trizac12,Samaj12}. If we focus on the one-particle density $\tilde{\rho}_1$, it is in fact exactly related to the fugacity $\lambda$ and the excess chemical potential $\mu_{ex}$ via the relation \cite{HansenMcDonald}:
\begin{equation} 
\tilde{\rho}_1(\mathbf{\tilde{r}}) = \lambda e^{-\phi(\mathbf{\tilde{r}})-\beta \mu_{ex}(\mathbf{\tilde{r}})} \label{eqPai21}
\end{equation}where $\phi$ is still the external potential in absence of any counterion. Contrary to the VSC approach that expands $\mu_{ex}$ in powers of the fugacity $\lambda$ (Mayer expansion), the aim here is to evaluate an exact expression valid for high values of $\Xi$ in the planar geometry: the many body problem is hence intrinsically accounted for in the WSC approach. One way to do it is to start from Widom's particle insertion method to compute the excess chemical potential \cite{Widom,Binder,FrenkelSmit}:
\begin{equation}
e^{-\beta \mu_{ex}(\mathbf{\tilde{r}})} = \langle e^{-\beta \Delta U}\rangle_{N-1} \label{eqPai22}
\end{equation}where $\Delta U$ is the variation of the whole interparticle energy when adding an Nth test particle at position $\mathbf{\tilde{r}}$ and $\langle .\rangle_{N-1}$ stands for a canonical average over all possible configurations of the N-1 other counterions.
\subsubsection{Case of one plate}
At high values of $\Xi$, the variation $\Delta U$ corresponds to a small perturbation of the interaction within a Wigner crystal formed by $N$ counterions by moving the Nth one away at a distance $\tilde{z}$ from the plate. According to \cite{TrizacPRL11,Trizac11} it reads within an harmonic approximation in the $z$ direction:
\begin{equation}
\beta \Delta U = -\frac{\alpha}{\sqrt{\Xi}} \sum_{i=1}^{N-1} \frac{(\tilde{z}-\tilde{z}_i)^2}{(|\mathbf{R}-\mathbf{R}_i|/a)^3} + \mathcal{O}(\Xi^{-1}) \label{eqPai23}
\end{equation}where $\alpha = 3^{3/4}/(16 \pi^{3/2})$, $\mathbf{R}_i$ represents the position of the lattice site $i$ and $a$ is the lattice spacing. Note that the value of $|\tilde{z}-\tilde{z}_i|$ is totally unrelated to that of $|\mathbf{R}-\mathbf{R}_i|$ and therefore, the canonical average reads\footnote{We use the fact that first:
\begin{eqnarray} 
\langle e^{-\beta \Delta U}\rangle_{N-1} = \langle 1 - \beta \Delta U + \mathcal{O}(\Xi^{-1}) \rangle_{N-1} \nonumber
\end{eqnarray}and second:
\begin{eqnarray}
 \langle \sum_{i=1}^{N-1}\frac{(\tilde{z}-\tilde{z}_i)^2}{(|\mathbf{R}-\mathbf{R}_i|/a)^3} \rangle_{N-1} =  \sum_{i=1}^{N-1}\frac{\langle(\tilde{z}-\tilde{z}_i)^2 \rangle_{N-1}}{(|\mathbf{R}-\mathbf{R}_i|/a)^3} = S \int_0^{\infty} d\tilde{z}'\:e^{-\tilde{z}'}(\tilde{z}-\tilde{z}')^2 + \mathcal{O}(\Xi^{-1/2}) \nonumber
\end{eqnarray}This means that the average $\langle (\tilde{z}-\tilde{z}_i)^2\rangle_{N-1}$ is dominated by the cost it takes to take the $i$th ion away from the plate. Other contributions to the average are of order $\Xi^{-1/2}$.}:
\begin{eqnarray} \langle e^{-\beta \Delta U}\rangle_{N-1} &=&  1 + \frac{\alpha S}{\sqrt{\Xi}}\int_0^{\infty} d\tilde{z}'\:e^{-\tilde{z}'}(\tilde{z}-\tilde{z}')^2 + \mathcal{O}(\Xi^{-1}) \label{eqPai25}
\end{eqnarray}
where $S = \sum_i 1/(|\mathbf{R}-\mathbf{R}_i|/a)^3 \approx 11.034$. Evaluating the integral in \eqref{eqPai25} and plugging it back into Eq. \eqref{eqPai21} we get:
\begin{equation}
\tilde{\rho}_1(\tilde{z}) = \frac{e^{-\tilde{z}}}{2\pi \Xi}\left[1 + \frac{3^{3/4}S}{ (4\pi)^{3/2}\sqrt{\Xi}}\left(\frac{\tilde{z}^2}{2}-\tilde{z}\right)  + \mathcal{O}(\Xi^{-1}) \right] \label{eqPai27}
\end{equation}where the prefactor solves the electroneutrality condition $\int dz \:q\rho_1 =2|\sigma|$. In eq. \eqref{eqPai27}, we see that corrections to the single particle picture are of order $\Xi^{-1/2}$ ---instead of the order $\Xi^{-1}$ prescribed in the VSC theory--- and agree very well with existing simulation data \cite{TrizacPRL11,Trizac11}. This discrepency weakens VSC approach's reliability beyond the single particle for planar geometries and makes its applicability uncertain for other geometries.
\subsection{Case of like charged plates}
The expansion for the density can be carried out in a similar way as for one plate explained above. It is not as easy as for one plate because the ground state itself depends on the distance $L$ between the plates \cite{Samaj12} and therefore we won't treat it fully in those lines. Three distance regimes can however be discriminated in a WSC approach and we shall see how they phenomenologically differ.
\subsubsection{Short distances} 
At short distances i.e. when $L \ll a_{\perp}$, the single particle picture can apply (i.e. $\Delta U$ in Eq. \eqref{eqPai22} is neglected at the single particle level) and yields an unbinding mechanism where the counterions detach from the plates\footnote{The typical lateral distance $a_{\perp}$ used here is a bit tricky since it depends on the distance between the plates \cite{Samaj12}. Let us say here that for small enough distances $L$, $a_{\perp}$ is defined as $2\pi a_{\perp}^2 |\sigma|= q $ while for infinite distances we have $\pi a_{\perp}^2 |\sigma|= q $.} \cite{Moreira01,Trizac12}. The particle density $\rho_1(z)$ is uniform and satisfies $\rho_1(z)=C e^{-\phi(z)}$, where $\phi(z) = cte$ for two equally charged plates and where $C$ is a constant too. Unsuring the electroneutrality condition \eqref{eqPai3}, the density within the slab reads $\rho_1(z)=2|\sigma|/qL$. Now, it is convenient to use the contact theorem that relates exactly the pressure $\beta P$ to the particle density at contact $\rho_1(0)$ \cite{Henderson,Wennerstrom}:
\begin{equation}
 \beta P = \rho_1(0) - 2\pi l_B \sigma^2 = \frac{2 |\sigma|}{qL}-2\pi l_B \sigma^2 \label{eqPai28}
\end{equation}This expression can lead to attraction if $\tilde{L}=L/\mu_G > 2$. 
\subsubsection{Intermediate distances}
This distance regime corresponds to separations $L$ such that $a_{\perp} < L < q^2l_B$. The unbinding mechanism responsible for a strong attractive pressure for short distances is not effective anymore and counterions start separating into two strongly correlated layers as $L$ increases \cite{Samaj12}. The density at the plate increases accordingly toward the zero pressure value $2 \pi l_B \sigma^2$. At large enough distance $\rho_1(0) \sim 2 \pi l_B \sigma^2 - f(L)$ where $f(L)$ is a small correction for large $L$. It can be shown that in the limit of infinite $\Xi$, $f(L)\sim e^{-G_0 L}$ where the characteristic length $G_0^{-1}$ depends on the Wigner crystal structure on the plates at large $L$ \cite{Samaj12}. This gives rise to an exponentially decaying attraction \cite{ShklovskiiPRL99,Lauthesis,Lau01,Samaj12} similar in spirit to the one found in the planar KL theory described above.
\subsubsection{Large distances}
If $L \gg q^2 l_B$, then counterions on a plate do not create anymore a proper correlation hole on the other plate and the condensed ionic layers are no longer strongly correlated with each other. Although the contact theorem \eqref{eqPai28} is valid for any $L$, it does not provide a simple understanding of the large distance physics. A simple way to understand it is to note that the two plates are so far away from each other that they are almost isolated and can be seen as dressed plates displaying a low charge density\footnote{Roughly speacking, the effective charge density $\sigma_{eff}$ scales as $\sim\:e^{-\sqrt{\Xi}}$ and goes rapidly to zero as the coupling parameter goes to infinity \cite{dosSantos,Paillusson-Trizac}.}. The counterion density far away from them is then described by a PB regime \cite{Burak04,dosSantos,Paillusson-Trizac}. At that point, the condensed counterion layers become an intrinsic property of these dressed plates akin to surface plasmons on facing metallic plates. Still, they remain correlated through charge fluctuations and yield a universal attraction at finite temperature \cite{Attard88,LauPRL2000, Lauthesis,Martin05}:
\begin{equation}
 \beta P_{att} = -\frac{\zeta(3)}{8\pi L^3} \label{eqPai29}
\end{equation}where $\zeta(x)$ is the Riemann zeta function and $\zeta(3) \approx 1.2021$. For Eq. \eqref{eqPai29} to give a net attractive pressure, it has to be compared with the PB repulsion owing to the counterion-dressed plates\footnote{The physics becomes very similar to the DLVO theory \cite{DLVO} where ionic contributions are accounted for by the PB theory and separated from the medium-related van der Waals interactions.}.

\section{Dumbbell like counterions}
\label{dumbbells_Paillusson}
\subsection{The model}
Among counterions of valency $n > 1$, some are molecules made of $n$ repetitions of a single charged unit (principally monovalent) \cite{Bohinc04,Bohinc09,Bohinc11,Bohinc12,Kim08,Kanduc09}. For the sake of illustration, we will focus on divalent positive conterions that neutralize two like-charged plates in absence of salt. The simplest description for a divalent counterion would be two monovalent point like ions of the same sign forming a dumbbell of length $l$.\newline Let us denote $p(\mathbf{r}, \mathbf{\hat{u}})$ the joint probability density to find a reference charge belonging to a dumbbell at $\mathbf{r}$ and having the dumbbell direction vector arising from it in direction $\mathbf{\hat{u}}$. The probability to find a dumbbell through its reference charge at position $\mathbf{r}$ reads:
\begin{equation}
 p(\mathbf{r}) \equiv \int d\Omega\: p(\mathbf{r}, \mathbf{\hat{u}}) \label{rod1}
\end{equation}where $d\Omega$ is the solid angle measure.

\subsection{Mean field theory}
\subsubsection{A modified PB equation}
Within a mean field theory, the joint probability density $p(\mathbf{r}, \mathbf{\hat{u}})$ for one dumbbell depends on the mean external electrostatic potential $\varphi$ generated by both its co-ions and the plates \cite{Bohinc04,Bohinc09,Kim08}:
\begin{equation}
 p(\mathbf{r}, \mathbf{\hat{u}}) \propto e^{-\beta e \varphi(z)}e^{-\beta e \varphi(z+l \cos(\theta))}H(z+l \cos \theta)H(L-(z+l \cos \theta)) \label{rod2}
\end{equation}where $\cos \theta $ is the projection of $\mathbf{\hat{u}}$ on the z-axis and $H(x)$ is the Heaviside step function that is zero if $x$ is negative and one otherwise. The Heaviside functions in Eq. \eqref{rod2} prevent possible overlaps between the dumbbell and the plates. The volume density of dumbbells $n(z)$ is proportional to the probability \eqref{rod1} and the corresponding charge density is $2e n(z)$ simply because there are two charges per dumbbell. Finally, for the definition of $\varphi$ to be consistent, it has to satisfy the Poisson equation (in I.S. units) which yields the modified Poisson-Boltzmann equation for dumbbells counterions in a slit geometry:
\begin{equation}
 \frac{d^2 \varphi}{dz^2} = -2e n(z)/\varepsilon = -\frac{eC}{\varepsilon l}\int_{-l_{min}(z)}^{l_{max}(z)}\:ds\: e^{-\beta e \varphi(z)}e^{-\beta e \varphi(z+s)} \label{rod4}
\end{equation}together with the boundary conditions:
\begin{equation}
 \left.\frac{d \varphi}{dz}\right|_{z=0}=\frac{e|\sigma|}{\varepsilon};\:\:\:\:\:\:\:\:\:\:\:\:\left.\frac{d \varphi}{dz}\right|_{z=L}=-\frac{e|\sigma|}{\varepsilon} \label{rod5}
\end{equation}that ensure the global electroneutrality of the system. In Eq. \eqref{rod4}, the change of variable $s = l \cos \theta$ has been made, the Heaviside measure appearing in Eq. \eqref{rod2} is unity in the interval $[l_{min}(z),l_{max}(z)]$ and $C$ is a constant of arbitrary value\footnote{To be physically consistent, the r.h.s. of Eq. \eqref{rod4} has to be invariant under the (gauge) transformation $\varphi \rightarrow \varphi +g$ where $g$ is a constant. This is only possible if, under this transformation, the constant $C$ undergoes a change $C \rightarrow Ce^{2g}$. Choosing a specific value for $C$ is thus equivalent to fixing the gauge $g$ and any value can then be taken for it \cite{Tamashiro03}.
}.
\subsubsection{Plate-plate interaction}
As before, we make use of the contact theorem that reads as follows for dumbbells \cite{Kim08,Bohinc09}:
\begin{equation}
 \beta P = 2 n(0) -2\pi l_B \sigma^2 \label{rod7}
\end{equation}The factor two in the osmotic part of Eq. \eqref{rod7} appears because there are two ends per dumbbell interacting with the plate at $z=0$. The solution to Eqs. \eqref{rod4} and \eqref{rod5} is a inhomogeneous potential $\varphi(z)$ that increases rapidly from a low value at the plate to its highest value at the mid-plane \cite{Bohinc09}. Two distance regimes can then be discriminated \cite{Kim08}. First, when the mid-interplate distance $L/2$ is larger than the size $l$ of the dumbbells. In this case, the rapid decrease of the potential $\varphi$ makes adsorbed counterions on a plate be oriented mostly parralel to it since it is very favorable energetically. Second, when $L/2$ is smaller than $l$, then adsorbed dumbbells on a plate can also lie perpendicularly to it by reaching a not so unfavorable potential close to the other plate \cite{Bohinc09}. The increase of such bridging configurations when $L > l$ implies that, on average, there will be less point charges trapped directly at the wall hence yielding a decrease in $n(0)$\footnote{If bridging configurations do exist then the dumbbell density will display four picks --- one at each plate and two others at a distance $l$ from the plates --- instead of simply two at the plates. Let $n_>(0)$ be the density at a plate when $L>2l$. The total number of dumbbells $N_d$ is mostly dominated by the two picks at the plates and scales roughly as $N_d \sim 2 n_>(0)$. Now, let $n_<(0)$ be the dumbbell density at a plate when $L < 2l$. It can be divided into parallel $n_p(0)$ and bridging $n_b(0)$ densities. In particular $n_b(0) = n_b(l)$ by definition of a bridge. The integral of the dumbbell density is now dominated by the four picks enumerated above and the total number of particles scales as $N_d \sim 2n_<(0) + 2n_b(0)$. It then turns out that as long as $n_b(0)$ is non zero and $L>l$, $n_<(0) < n_>(0)$.}. This drop can be sufficient enough for the electrostatic attraction to overcome the osmotic pressure and then generate an attraction. Now, if $L\leq l$, there is no lack of point charges at the plates and the interaction is repulsive again \cite{Kim08,Bohinc09}. This allows then to bound the equilibrium distance owing to a PB bridging mechanism $l<L_{eq}<2l$.
\subsection{SC regime for dumbbells}
If the coupling parameter $\Xi$ is big enough then a SC description can be done on dumbbell counterions neutralizing two plates \cite{Kim08,Bohinc12}. In particular, if $l < a_{\perp}$, then a single particle picture can be used. In this case the dumbbell density reads:
\begin{equation}
 n(z) = \frac{\Omega(z)}{4\pi}K e^{- \tilde{L}} \label{rod8}
\end{equation}where $\Omega(z)$ is the accessible solid angle for the dumbbell at location $z$, $\tilde{L} = L 2\pi l_B |\sigma|q$ (with $q=2$) is the constant external potential felt by the dumbbell and $K$ is a normalization constant. The constant $K$ is chosen to satisfy electroneutrality and reads $K = |\sigma|e^{\tilde{L}}/(L-l/2)$ when $l<L$\footnote{If $L>l$, the factor $\Omega(z)/4\pi$ is $1/2+z/(2l)$ if $0<z<l$, $1/2+L/2-z/(2l)$ if $L-l<z<L$ and unity otherwise. Its integrated value over the whole slab is thus simply $L-l/2$.}. We can then apply the contact theorem \eqref{rod7} which yields:
\begin{equation}
 \beta P = \frac{|\sigma|}{(L-l/2)} -2\pi l_B \sigma^2 \label{rod9}
\end{equation}Hence a stable bridging equilibrium arises at $\tilde{L}_{eq} = l+2$. Likewise, when $l>L$, the accessible solid angle is $\Omega(z)=2\pi L/l$ for any $z$ \cite{Kim08} and the electroneutrality condition yields $n(z) = |\sigma|/L$. Inserting back this expression into the contact theorem \eqref{rod7}, we get a stable equilibrium $\tilde{L}_{eq} = 4$ that is independent of the dumbbell size. Note also that it is twice the equilibrium length that would be obtained for point-like divalent ions in the SC regime in \eqref{eqPai20}. This is because the attraction is here due to electrostatic correlations (and not a bridging mechanism) that are weaker with a divalent dumbbell than with a concentrated charge.

\subsection{Validity domain and the point-like limit}
If $l$ is small enough, we should recover the physics of point like divalent ions. This happens when the dumbbell orientation is unaffected by the electrostatic potential gradient i.e. when $l \ll \mu_G < L $ \cite{Kim08,Bohinc09}. In this case, $n(z)$ becomes $n(z)= \rho_1(z)\Omega(z)/4\pi$ where $\rho_1(z)$ is the point-like divalent counterion density and $\Omega(z)$ is the accessible solid angle for the dumbbell. In particular, however small the dumbbell is, at $z=0$ only half of the total solid angle is accessible to the dumbbell and this allows us to recover the contact theorem for point-like ions \eqref{eqPai28} from the contact theorem for dumbbells Eq. \eqref{rod7}.
\section{DNA-DNA attraction}
 It is now well established that molecular (e.g. polyamines) and atomic (e.g. Cr$^{+3}$) cations can lead to the formation of hexagonal arrays of dsDNA provided their valency is higher than $+2$ even though some divalent ions (e.g. Mn$^{+2}$ and Cd$^{+2}$) are able to condense dsDNA \cite{bloomfield96,Teif11,Kornyshev07}. Such arrays have been experimentally studied recently \cite{Todd, DeRouchey} and both attractive and repulsive contributions to the interaction appear to be exponentially decaying. A long ranged attraction as seen in Eq. \eqref{eqPai29} may still exist but it does not seem to drive the array stability.\newline Todd et al. \cite{Todd} also found the attractive decay length $\lambda_{att}$ to be almost indepedent of the counterion type ($\lambda_{att} \sim 5 \AA$) and to be roughly twice the repulsive characteristic length $\lambda_{rep}$. It seems  unlikely for a bridging mechanism to be compatible with these observations as its physics depends very much on the length of the molecular cations (see section \ref{dumbbells_Paillusson}.). Moreover, the corresponding equilibrium distance $L_{eq}$ should be about the size $l$ of the chains and thus should be increasingly big as the valency of the cations increases. This is in contradiction with what is observed in experiments \cite{Todd, DeRouchey} where the DNA-DNA equilibrium distance decreases while increasing the counterion valency. The KL theory and the WSC approach however, are both compatible with an exponentially decaying attraction. Even though the WSC has yet to be fully treated for a realistic dsDNA, it is likely for the length $G_0^{-1}$ (see section \ref{SC_Paillusson}) to roughly equal dsDNA's mean helical pitch $H \approx 3.4 \:\rm nm$ \cite{ShklovskiiPRL99}. This would then lead to an attractive decay length $\lambda_{att}\approx H/2\pi \approx 5.4 \:\AA$ (that is not so far from $\lambda_{att}$) as prescribed in the original KL theory \cite{Kornyshev99,Kornyshev01}. In addition, the ratio between DNA and water electrostatic permitivities being about $1/80$, it generates an exponentially decaying repulsion --- owing to image charges of the dressed dsDNAs within each facing dsDNA \cite{Jackson} ---  with the same decay length as for the attraction but for a distance that is twice that of the spacing between the dsDNAs \cite{DeRouchey,Kornyshev97}. In the end, it is equivalent to a repulsion whose decay length is exactly half that of the attraction decay length as measured in studies \cite{Todd} and \cite{DeRouchey}.\newline The KL theory seems in very good agreement with experiments \cite{Kornyshev05,Kornyshev07,Todd,DeRouchey} although some of the assumptions it is based on arguable\cite{Granot82_1,Granot82_2,Granot82_3}. A SC analysis seems potentially compatible and could provide some rational basis to the effective parameters of the KL theory \cite{Kornyshev05,Kornyshev07,Kanduc08,DeRouchey,Qiu10}.

Finally, the above discussion on dsDNA condensation only accounts for the improvements of the PB theory mentioned in the introduction. An obvious refinement that has not been treated in these lines is a better model for water. Such a thing goes way beyond the scope of this chapter. It is however worth mentionning that a phenomenological account of the way polyvalent counterions could affect the structure of water on dsDNA can give rise to an attraction \cite{Parsegian85,Parsegian92} whose quantitative features agree well with the work of \cite{Todd,DeRouchey}.

\bibliographystyle{unsrt} 
\bibliography{Biblio_Contrib_Paillusson} 

\begin{thebibliography}{10}

\bibitem{Parsegian}
V.~A. Parsegian and D.~Gingell.
\newblock On the electrostatic interaction across a salt solution between two
  bodies bearing unequal charges.
\newblock {\em Biophys. J.}, 12:1192--1204, 1972.

\bibitem{Neu}
J.C. Neu.
\newblock Wall-mediated forces between like-charged bodies in an electrolyte.
\newblock {\em Phys.Rev.Lett.}, 82:1072, 1999.

\bibitem{Sader}
J.E. Sader and D.Y.C Chan.
\newblock Long range electrostatic attractions between identically charged
  particles in confined geometries: an unresolved problem.
\newblock {\em J.Coll.Int.Sci.}, 213:268--269, 1999.

\bibitem{Orland}
R.R. Netz and H.~Orland.
\newblock Beyond poisson-boltzmann: Fluctuations and correlations.
\newblock {\em Eur.Phys.J. E}, 1:203--214, 2000.

\bibitem{HL00}
J.-P. Hansen and H.~L\"owen.
\newblock {\em Annu. Rev. Phys. Chem.}, 51:209, 2000.

\bibitem{Kornyshev99}
A.A. Kornyshev and S.~Leikin.
\newblock Electrostatic zipper motif for dna aggregation.
\newblock {\em Phys.Rev.Lett.}, 82:4138--4141, 1999.

\bibitem{Kornyshev01}
A.A. Kornyshev and S.~Leikin.
\newblock Sequence recognition in the pairing of dna duplexes.
\newblock {\em Phys.Rev.Lett.}, 86:3666, 2001.

\bibitem{Kornyshev07}
A.A. Kornyshev, D.~J. Lee, S.~Leikin, and A.~Wynveen.
\newblock Structure and interactions of biological helices.
\newblock {\em Rev.Mod.Phys.}, 79:943, 2007.

\bibitem{Kornyshev11}
R.~Cortini, A.A. Kornyshev, D.J. Lee, and S.~Leikin.
\newblock Electrostatic braiding and homologous pairing of dna double helices.
\newblock {\em Biophys.J.}, 101:875--884, 2011.

\bibitem{Rouzina96}
I.~Rouzina and V.A. Bloomfield.
\newblock Macroion attraction due to electrostatic correlation between
  screening counterions. 1. mobile surface-adsorbed ions and diffuse ion cloud.
\newblock {\em J. Phys. Chem.}, 100:9977, 1996.

\bibitem{ShklovskiiPRL99}
B.I. Shklovskii.
\newblock Wigner crystal model of counterion induced bundle formation of
  rodlike polyelectrolytes.
\newblock {\em Phys.Rev.Lett.}, 82:3268, 1999.

\bibitem{Lau01}
A.W-C. Lau, P.~Pincus, and H.A. Levine, D.~Fertig.
\newblock Electrostatic attraction of coupled wigner crystals: finite
  temperature effects.
\newblock {\em Phys.Rev.E}, 63:051604, 2001.

\bibitem{Lauthesis}
A.~W-C. Lau.
\newblock {\em Fluctuation and Correlation effects in Electrostatics of
  Highly-Charged Surfaces (PhD thesis)}.
\newblock 2000.

\bibitem{Moreira02}
A~Moreira and R.R. Netz.
\newblock Simulations of counterions at charged plates.
\newblock {\em Eur. Phys. J. E}, 8:33, 2002.

\bibitem{HansenMcDonald}
J.-P. Hansen and I.R. McDonald.
\newblock {\em Theory of simple liquids (third edition)}.
\newblock Academic Press, 2006.

\bibitem{Netz01}
R.R. Netz.
\newblock Electrostatistics of counter-ions at and between planar charged
  walls: From poisson-boltzmann to the strong-coupling theory.
\newblock {\em Eur. Phys. J. E}, 5:557, 2001.

\bibitem{Moreira01}
A.G. Moreira and R.~Netz.
\newblock Field-theoretic approaches to classical charged systems.
\newblock In {\em Electrostatic Effects in Soft Matter and Biophysics (C. Holm,
  P. K\'ekicheff and R Podgornik, Eds; NATO Science Series)}, pages 367--408.
  Kluwer Academic Publishers, 2001.

\bibitem{Netz04}
A.~Naji and R.R. Netz.
\newblock Attraction of like-charged macroions in the strong-coupling limit.
\newblock {\em Eur.Phys.J.E}, 13:43, 2004.

\bibitem{Naji04}
A.~Naji, A.~Arnold, C.~Holm, and R.R. Netz.
\newblock Attraction and unbinding of like-charged rods.
\newblock {\em Europhys.Lett.}, 67:130, 2004.

\bibitem{KanducPRE08}
M.~Kandu\ifmmode~\check{c}\else \v{c}\fi{}, M.~Trulsson, A.~Naji, Y.~Burak,
  J.~Forsman, and R.~Podgornik.
\newblock Weak- and strong-coupling electrostatic interactions between
  asymmetrically charged planar surfaces.
\newblock {\em Phys. Rev. E}, 78(6):061105, Dec 2008.

\bibitem{TrizacPRL11}
Ladislav \ifmmode~\check{S}\else \v{S}\fi{}amaj and Emmanuel Trizac.
\newblock Counterions at highly charged interfaces: From one plate to
  like-charge attraction.
\newblock {\em Phys. Rev. Lett.}, 106(7):078301, 2011.

\bibitem{Trizac11}
L.~Samaj and E.~Trizac.
\newblock Wigner-crystal formulation of strong-coupling theory for counterions
  near planar charged interfaces.
\newblock {\em Phys.Rev.E}, 84:041401, 2011.

\bibitem{Shklovskii99}
B.~I. Shklovskii.
\newblock Screening of a macroion by multivalent ions: Correlation-induced
  inversion of charge.
\newblock {\em Phys. Rev. E}, 60(5):5802--5811, Nov 1999.

\bibitem{LauPRL2000}
A.W-C. Lau, D.~Levine, and P.~Pincus.
\newblock Novel electrostatic attraction from plasmon fluctuations.
\newblock {\em Phys.Rev.Lett.}, 84:4116, 2000.

\bibitem{Paillusson-Trizac}
F.~Paillusson and E.~Trizac.
\newblock Interaction regimes for oppositely charged plates with multivalent
  counterions.
\newblock {\em Phys. Rev. E}, 82:052501, 2011.

\bibitem{Trizac12}
L.~Samaj and E.~Trizac.
\newblock Strong-coupling theory for a polarizable planar colloid.
\newblock {\em Contributions to Plasma Physics}, 52:53, 2012.

\bibitem{Samaj12}
L.~Samaj and E.~Trizac.
\newblock Critical phenomena and phase sequence in classical bilayer wigner
  crystal at zero temperature.
\newblock {\em Phys.Rev.B}, 85:205131, 2012.

\bibitem{Widom}
B.~Widom.
\newblock Some topics in the theory of fluids.
\newblock {\em J.Chem.Phys.}, 39:2802, 1963.

\bibitem{Binder}
K.~Binder.
\newblock Applications of monte carlo methods to statistical physics.
\newblock {\em Rep.Prog.Phys.}, 60:487, 1997.

\bibitem{FrenkelSmit}
D.~Frenkel and Smit B.
\newblock {\em Understanding molecular simulation (second edition)}.
\newblock Academic Press, 2002.

\bibitem{Henderson}
D.~Henderson and L.~Blum.
\newblock {\em J. Chem. Phys.}, 69:5441, 1978.

\bibitem{Wennerstrom}
H.~Wennerstr\"om, B.~J\"onsson, and P.~Linse.
\newblock The cell model for polyelectrolyte systems. exact statistical
  mechanical relations, monte carlo simulations, and the poisson-boltzmann
  approximation.
\newblock {\em J. Chem. Phys.}, 76:4665--4670, 1982.

\bibitem{dosSantos}
A.P. dos Santos, A.~Diehl, and Yan Levin.
\newblock Electrostatic correlations in colloidal suspensions: Density profiles
  and effective charges beyond the poisson-boltzmann theory.
\newblock {\em J. Chem. Phys.}, 130:124110, 2009.

\bibitem{Burak04}
Y.~Burak, D.~Andelman, and H.~Orland.
\newblock Test-charge theory for the electric double layer.
\newblock {\em Phys. Rev. E}, 70:016102, 2004.

\bibitem{Attard88}
P.~Attard, R.~Kjellander, D.J. Mitchell, and B.~J\"onson.
\newblock Electrostatic fluctuation interactions between neutral surfaces with
  adsorbed, mobile ions or dipoles.
\newblock {\em J.Chem.Phys.}, 89:1664, 1988.

\bibitem{Martin05}
P.R. Buenzli and Ph.A. Martin.
\newblock The casimir force at high temperature.
\newblock {\em Europhys.Lett.}, 72:42--48, 2005.

\bibitem{DLVO}
J~Vervey and J.~T.~G. Overbeek.
\newblock {\em Theory of the Stability of Lyophobic collo\"ids}.
\newblock Elsevier: Amsterdam, 1948.

\bibitem{Bohinc04}
K.~Bohinc, A.~Igli\v{c}, and S.~May.
\newblock Interaction between macroions mediated by divalent rod-like ions.
\newblock {\em Europhys.Lett.}, 68:494, 2004.

\bibitem{Bohinc09}
S.~Maset, J.~Re\v{s}\v{c}i\v{c}, S.~May, J.I. Pavli\v{c}, and K.~Bohinc.
\newblock Attraction between like-charged surfaces induced by orientational
  ordering of divalent rigid rod-like counterions: theory and simulations.
\newblock {\em J.Phys.A:Math.Theor.}, 42:105401, 2009.

\bibitem{Bohinc11}
K.~Bohinc, J.~Re\v{s}\v{c}i\v{c}, S.~Maset, , and S.~May.
\newblock Debye-hückel theory for mixtures of rigid rodlike ions and salt.
\newblock {\em J.Chem.Phys.}, 134:074111, 2011.

\bibitem{Bohinc12}
K.~Bohinc, J.M.A. Grime, and L.~Lue.
\newblock The interactions between charged colloids with rod-like counterions.
\newblock {\em Soft Matter}, 8:5679, 2012.

\bibitem{Kim08}
Y.W. Kim, J.~Yi, and P.A. Pincus.
\newblock Attractions between like-charged surfaces with dumbbell-shaped
  counterions.
\newblock {\em Phys.Rev.Lett.}, 101:208305, 2008.

\bibitem{Kanduc09}
M.~Kandu\v{c}, A.~Naji, Y.S. Jho, P.A. Pincus, and R.~Podgornik.
\newblock Role of multipoles in counterion-mediated interactions between
  charged surfaces: Strong and weak coupling.
\newblock {\em J.Phys.: Cond. Matt.}, 21:424103, 2009.

\bibitem{Tamashiro03}
M.N. Tamashiro and H.~Schiessel.
\newblock Where the linearized poisson-boltzmann cell model fails: The planar
  case as a prototype study.
\newblock {\em Phys.Rev.E}, 68:066106, 2003.

\bibitem{bloomfield96}
V.A. Bloomfield.
\newblock Dna condensation.
\newblock {\em Curr.Opin.Struct.Biol.}, 6:334--341, 1996.

\bibitem{Teif11}
V.B. Teif and K.~Bohinc.
\newblock Condensed dna: Condensing the concepts.
\newblock {\em Prog. Biophys. Mol. Biol.}, 105:208, 2011.

\bibitem{Todd}
B.A. Todd, V.A Parsegian, A.~Shirahata, T.J. Thomas, and D.C. Rau.
\newblock Attractive forces between cation condensed dna double helices.
\newblock {\em Biophys. J.}, 94:4775, 2008.

\bibitem{DeRouchey}
J.~DeRouchey, Parsegian V.A., and D.C. Rau.
\newblock Cation charge dependence of the forces driving dna assembly.
\newblock {\em Biophys.J.}, 99:2608--2615, 2010.

\bibitem{Jackson}
J.D. Jackson.
\newblock {\em Classical electrodynamics (third edition)}.
\newblock Wiley, 1998.

\bibitem{Kornyshev97}
A.A. Kornyshev and S.~Leikin.
\newblock Theory of interaction between helical molecules.
\newblock {\em J.Chem.Phys.}, 107:3656--3674, 1997.

\bibitem{Kornyshev05}
A.A. Kornyshev, D.J. Lee, S.~Leikin, Wynveen A., and S.~B. Zimmermann.
\newblock Direct observation of azimuthal correlations between dna in hydrated
  aggregates.
\newblock {\em Phys.Rev.Lett.}, 95:148102, 2005.

\bibitem{Granot82_1}
J.~Granot, J.~Feignon, and D.R. Kearns.
\newblock Interactions of dna with divalent metal ions. i. $^{31}$p-nmr
  studies.
\newblock {\em Biopolymers}, 21:181, 1982.

\bibitem{Granot82_2}
J.~Granot and D.R. Kearns.
\newblock Interactions of dna with divalent metal ions. ii. proton relaxation
  enhancement studies.
\newblock {\em Biopolymers}, 21:203, 1982.

\bibitem{Granot82_3}
J.~Granot and D.R. Kearns.
\newblock Interactions of dna with divalent metal ions. iii. extent of metal
  binding: experiment and theory.
\newblock {\em Biopolymers}, 21:219, 1982.

\bibitem{Kanduc08}
M.~Kanduc, J.~Dobnikar, and R.~Podgornik.
\newblock Counterion-mediated electrostatic interactions between helical
  molecules.
\newblock {\em Sof matter}, 5:868--877, 2009.

\bibitem{Qiu10}
X~Qiu, V.A. Parsegian, and D.C. Rau.
\newblock Divalent counterion-induced condensation of triple-strand dna.
\newblock {\em PNAS}, 107:21482?21486, 2010.

\bibitem{Parsegian85}
V.A. Parsegian, R.P. Rand, and D.C. Rau.
\newblock Hydration forces: what next?
\newblock {\em Chem.Scr.}, 25:28--31, 1985.

\bibitem{Parsegian92}
D.C. Rau and V.A. Parsegian.
\newblock Direct measurement of the intermolecular forces between
  counterion-condensed dna double helices: evidence for long range attractive
  hydration forces.
\newblock {\em Biophys.J.}, 61:246--259, 1992.

\end{thebibliography}

\end{document}